\documentclass[pra,twocolumn,showpacs]{revtex4}
\usepackage{graphicx}
\usepackage{amsmath}
\usepackage{natbib}
\usepackage{epsfig}

\newcommand{\n}{\nonumber}
\newcommand{\bn}{\begin{eqnarray}}
\newcommand{\en}{\end{eqnarray}}
\newcommand{\eml}{\end{multline}}
\newcommand{\bml}{\begin{multline}}
\newcommand{\h}{\hspace}

\begin{document}

\title {Transparent, Non-local, Species-selective Transport in an Optical Superlattice Containing Two Interacting Atom Species}
 \author{Miroslav Gajdacz$^1$, Tom\'{a}\v{s} Opatrn\'{y}$^1$, and Kunal K. Das$^2$ }
 \affiliation{
 $^1$Optics Department, Faculty of Science, Palack\'{y} University, 17.  Listopadu 12,
 77146 Olomouc, Czech Republic\\
 $^2$Department of Physical Sciences, Kutztown University of Pennsylvania, Kutztown, Pennsylvania 19530, USA}

\date{\today }
\begin{abstract}
In an optical superlattice of triple wells, containing two mutually interacting atom species in adjacent wells, we show that one species can be transported through the positions of the other species, yet avoiding significant overlap and direct interaction. The transfer protocol is optimized to be robust against missing atoms of either species in any lattice site, as well as against lattice fluctuations. The degree and the duration of the inter-species overlap during passage can be tuned, making possible controlled large-scale interaction-induced change of internal states.
\end{abstract}
\maketitle

\section{Introduction}
Optical superlattices offer the means to scale up quantum operations on individual units of ultracold atoms and molecules for parallel processing a multitude simultaneously. Tremendous progress has been made towards controlled manipulations in double well lattices \cite{Mandel-Nature2003, Phillips-PRL2007, Phillips-Nature2007} allowing targeted change of internal atomic states and controlled merger and separation of pairs of atoms or their internal states. In this paper, we propose a new level of control achievable in a triple-well superlattice; wherein two mutually interacting species can be transported through each other with very little (precisely controlled) overlap at any time.  The method dubbed CTAP (coherent transport adiabatic passage) operates by principles similar to that of STIRAP (stimulated Raman adiabatic passage, \cite{STIRAP-RMP}) - an atom in a triple-well (three-levels) is transferred between the extreme wells (states) avoiding the middle well (state). The phenomenon has attracted attention in the context of single species in stand-alone quantum wells \cite{Greentree,Eckert,Graefe,Mompart-PRA2010,Opatrny-Das-PRA2009,Nesterenko-2009}, but has been demonstrated only with photons in waveguides \cite{Longhi-PRB2007}. Setting aside its novelty value as an interesting quantum effect, our goal in this paper is to gauge its true applications potential by: (i) applying it to a lattice (L\emph{attice}-CTAP) with a view to scaling-up and parallel processing of atomic operations, (ii) considering two intermingled interacting species, with controllable overlap during spatial manipulation, and (iii) ensuring  transfer algorithms optimized to work \emph{with or without} a second species, to allow for lattice imperfections and for applications to entanglement. We map out effects of physical factors and their parametric dependencies, and show the feasibility of experiments.

Inasmuch as it involves dynamically controlled coherent transfer of a quantum state to a very specific target state,  LCTAP can be viewed in the context of optimal control of quantum systems, which has seen much recent interest in the context time-dependent quantum processes such as loading of Bose-Einstein condensates onto optical lattices \cite{Sklarz-PRA-2002}  and of quantum state  transport  along spin chains  \cite{Venuti-PRA-2007,Caneva-PRL-2009,Murphy-PRA-2010}.
As in such phenomena, LCTAP crucially relies on optimizing the evolution of quantum states of atoms in a superlattice, in achieving nonlocal transport that avoids obstructions in the classical path.

The paper is organized as follows: In Sec. \ref{sec-Model} we describe our physical model and present our primary results demonstrating successful LCTAP evolution in a superlattice. In Sec. \ref{sec-Analysis} we present our analysis for appropriate optimization of the system parameters to achieve LCTAP evolution, specifically highlighting the inherent compromise necessary among various desired features. We demonstrate feasibility of our results in Sec. \ref{sec-Parameters} by relating the parameters in our simulations to realistic physical parameters in current experiments. We conclude with comments in Sec. \ref{Conclusion} on the assumptions and outlook of our model. Details of our numerical simulations are provided in Appendix \ref{appA}.

\section{Model and Results}
\label{sec-Model}
\begin{figure}[t]
\includegraphics*[width=\columnwidth]{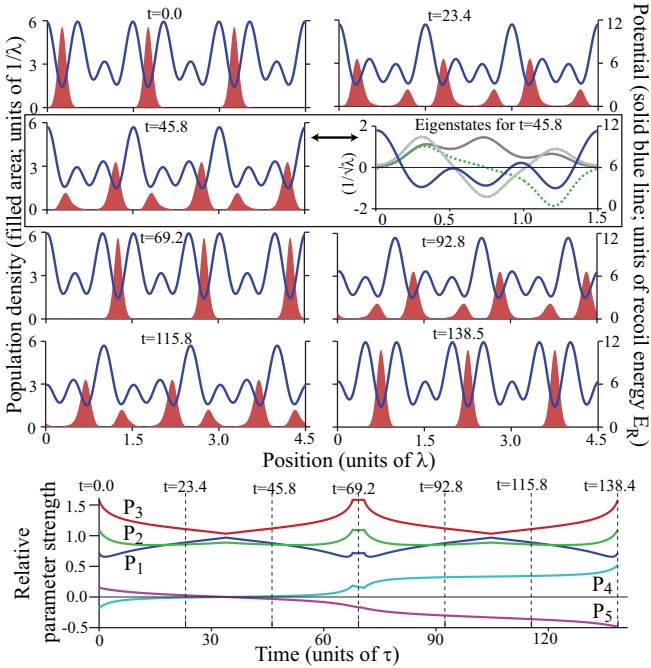}\vspace{-4mm}
\caption{(Color online) LCTAP of a single species:
Snapshots of population density (solid red, left axis) and
lattice potential (blue line, right axis) for
\emph{three} adjacent super-cells over \emph{two cycles}. The three
lowest eigenstates are shown for snapshot 3 (t=45.8 $\tau$). Snapshot times are marked on the
evolution of the amplitudes $P_1,P_2,P_3$, scaled by $4E_R$, and
relative phases $P_4,P_5$, scaled by $2\pi$. \emph{Movies of evolution are available online} \cite{movies}.
}\label{1-species-snapshots}\vspace{-5mm}
\end{figure}

We consider an optical
superlattice with unit cells of triple-wells, that could be
implemented with lasers of a single wavelength $\lambda$ at three
orientations \cite{Phillips-PRA2003} with shifted
frequencies. The lattice potential,
\bn
\label{potential}
V(x,t)=4E_R\left[P_1(t)\cos[2\pi({\textstyle\frac{2}{3}}x+P_4(t))]\h{1.5cm}\right.\n\\
\left.+P_2(t)\cos[2\pi({\textstyle\frac{4}{3}}x+P_5(t))]+P_3(t)\cos(4\pi
x)\right],\en is a superposition of periodicities $\lambda/2$,
$3\lambda/4$, $3\lambda/2$. The lattice depths $P_1,P_2,P_3$ (scaled
by $4\times$ recoil energy, $E_R=h^2/2m\lambda^2$ of target species
A) and the two relative phases $P_4,P_5$ (scaled by $2\pi$)
constitute five independent parameters to be varied in time for
LCTAP. We first examine and optimize LCTAP for a single \emph{target} species (A) in a superlattice before
generalizing for the presence of \emph{spectator} species (B).  An approach based on the lowest three eigenstates, ground (g), dark (d) and excited (e), of each unit triple-well cell, is suitable since adjacent unit cells are kept well isolated. LCTAP hinges on maintaining the system in the dark state \cite{Opatrny-Das-PRA2009}, since the dark state is antisymmetric with a node in the middle of the cell (as can be seen in one of the frames of Fig.~\ref{1-species-snapshots}), leading to minimal population in the central-well as compared to the other two states. Our numerical results assume length, energy
and time units of $\lambda,E_R,$ and $\tau = \hbar /E_R$. The time evolution was done by numerically propagating the Schr\"{o}dinger equation for the lattice potential Eq.~(\ref{potential}) using the split-step operator method based upon standard fast Fourier transform (FFT) routines.

\subsection{Single Species}
\begin{figure}[b]
\includegraphics*[width=\columnwidth]{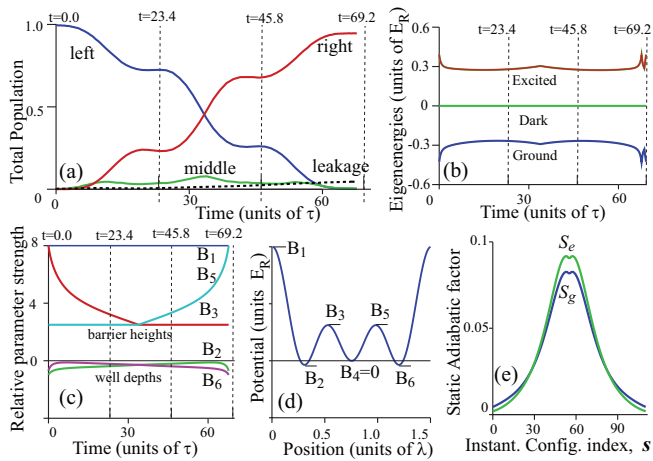}
\caption{(Color online) LCTAP for a single species corresponding to Fig.~\ref{1-species-snapshots} shown for
one cell and one cycle: (a) Evolution of the population
density in the three wells; the flat dotted line is the leakage out of the unit cell. (b) Evolution of the energies  with \emph{dark state as reference}; (c) Evolution of the barrier heights and well depths
indicated in (d). (e) Evolution of the rate-independent adiabatic
factors for the ground and excited states as a functions of index ($s$)
labeling the  instantaneous states.
}\label{1-species-parameters}
\end{figure}

In the case of LCTAP with a single species (A), the lattice is pattern-loaded with
one atom in the left-most well of each unit cell
\cite{Phillips-PRA2003,Bloch-PRL2008}. Our optimized LCTAP achieves simultaneous
transfer of all the atoms from well 1 (left) to well 3
(right) without significant occupation of well 2 (middle)
during the transfer. We demonstrate sustainability by
repeating transport from well 3 to well 5 (well 2 of the next cell)
avoiding well 4. Snapshots of the transport in
Fig.~\ref{1-species-snapshots} show atom density in three adjacent unit cells,
and the corresponding lattice potential. Low leakage out of each cell and low middle
well population at all times are  evident in plots Fig.~\ref{1-species-parameters}(a) of evolution of the atom density in a cell.

\begin{figure}[t]\vspace{-0.8cm}
\includegraphics*[width=\columnwidth]{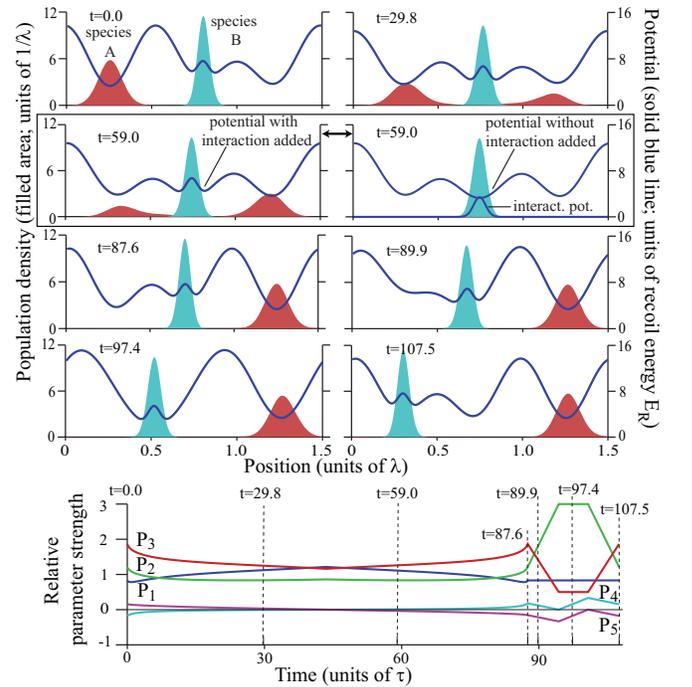}
\caption{(Color online) LCTAP with two species:
Snapshots \cite{movies} of population density (left axis; broad solid red: species A;
narrow solid cyan: species B) and
lattice potential $V_A$ (blue line, right axis) shown for \emph{one}
unit cell and \emph{one cycle}. The effect of A-B interaction
appears as a `bump' in the potential seen by A; shown separated out for snapshot 3. Evolution of lattice parameters $P_i$ \emph{is the same with or without the atom B present}. }\label{2-species-snapshots}
\end{figure}

\subsection{Dual Species}
Each unit cell is loaded with an atom A in well 1 and
an atom B in well 2. Species-selective trapping
\cite{Thywissen-PRA2007,Dominik-Schneble-PRL2010} is used to achieve
much deeper trap-depth for species B ($V_B/V_A\sim 10$), to keep its
density profile unaffected by lattice variations during LCTAP of
species A, which in evident in the snapshots in
Fig.~\ref{2-species-snapshots}.

The static shape of the atom B profile justifies a single particle
approximation, treating the A-B interaction as an added external
potential felt by atom A. We model it on the 1D bosonic hard-core
interaction $g_{1D}\delta(x_A-x_B)$ where $g_{1D}=2a\hbar\omega_\perp$ for a tight
cylindrical transverse harmonic confinement \cite{das-crossover} of frequency
$\omega_\perp$, scattering length $a$. Allowing
for the density spread, we replace the delta function with the
ground state density of atom B in the middle well,
$\delta(x_A-x_B)\rightarrow |\psi^B_0(x)|^2$, so the effect of the
interaction on A is modeled as a potential barrier (for positive
$a$) with position and width of $|\psi^B_0(x)|^2$ and strength
determined by $g_{1D}|\psi^B_0(x_B)|^2$. The barrier appears as a `bump'
in the potential (Fig.~\ref{2-species-snapshots}) seen by A.

The first four snapshots in Fig.~\ref{2-species-snapshots} demonstrate LCTAP of atom A from well 1 to well 3
avoiding overlap with atom B in well 2 during the evolution. In the last three snapshots atom B is transported oppositely to restore initial configuration (shifted), for the next cycle - also achievable by manipulating A alone
leaving B in place, but requires more complex
parameter evolution. The crucial element of this transfer is illustrated in
Figs.~\ref{2-species-parameters}(a) and
(b), showing similar LCTAP of species A \emph{with} or \emph{without} the species B actually present, with the same low tolerances on middle well population of A and leakage, and  most importantly, achieved for both cases \emph{with exactly the same parameter variation} seen in Fig.~\ref{2-species-snapshots}.

The results discussed here are based on the approximation of tight binding of atoms B in the lattice. The approximation is reasonable provided that the interaction energy between atoms A and B is much smaller than the energy necessary to excite atom B to a higher vibrational state. If this approximation is relaxed, the dynamics would contain entanglement of the atomic spatial variables during the transport, and the dimensionality of the system Hilbert space would increase. This more complex scenario can have interesting implications that will be explored elsewhere, but is not necessary for the basic two species LCTAP mechanism we study here, using species-specific lattice potential to provide the necessary tight confinement of atoms B.

\section{Analysis and Implementation}
\label{sec-Analysis}
LCTAP
is made challenging by the fact that there are more physical features to be optimized than the five
degrees of freedom offered by the lattice parameters. Here we identify the influence of those features
and their parametric dependencies, to achieve: (i) minimal middle well
population at all times (ii) minimal tunneling out of each triple
well cell (iii) maximal transfer into third well per cycle (iv)
robustness against noise and (v) maximal transfer speed.

Effective LCTAP hinges on the non-adiabaticity
factor
\bn {\cal A}_j(\psi_j,\psi_d)=\frac{|\langle
\psi_j|\textstyle\frac{\partial V}{\partial
t}|\psi_d\rangle|}{(E_j-E_d)^2}\h{1cm} j=g,e\label{adiabfactor}\en
defined by instantaneous states and their energies.  Two distinct
dependencies are manifest: (i) the energy separations $\Delta
E_i=E_i-E_d$ in the denominator, independent of, and (ii) the numerator, linearly dependent
on the rate of the change of the Hamiltonian. The time
rate of change can therefore be factored out $ {\cal A}_j={\cal
S}_j\frac{ds}{dt}$, with $s$ being a configuration index labelling the
instantaneous states.  Thus, the optimization is done in
two steps: (i) one finds the optimal sequence of \emph{static}
instantaneous states and (ii) optimizes the \emph{dynamic}
transfer rate factor along that path.


\subsection{Setting Dynamic Conditions}
An intuitive connection to the triple-well case can be made by reparameterizing the system  in terms of local parameters, such as the set of barrier heights and well depths $\{B_i,i\in[1,2,3,5,6]\}$ of individual triple-well unit cells, shown in Fig.~\ref{1-species-parameters}(d), where the middle well depth, $B_4$, serves as reference. Within our localized state approximation, the path for optimal LCTAP evolution can be defined precisely in terms of these local parameters.  Then the corresponding optimal path of the lattice parameters $\{P_i,i\in[1,5]\}$ is determined by an iterative numerical procedure, which is described in detail in Appendix \ref{appA}.  The evolution of the local parameters is directly dictated by our optimization conditions discussed below, and are shown in Fig.~\ref{1-species-parameters}(c) for the case of single species. The corresponding evolution of the original parameters, shown in Fig.~\ref{1-species-snapshots}, are numerically computed from these using the prescription in  Appendix \ref{appA}.

By means of this reparameterization, determining the optimal LCTAP path translates to finding the optimal evolution of an  appropriate set of five local parameters, that can be mapped to the original five lattice parameters. As with STIRAP, minimally two time-varying parameters are necessary for the transfer. This is accomplished by counter-intuitive variation of the internal barrier heights $B_3$ and $B_5$ that fix the couplings of the wells, as can be seen in Fig.~\ref{1-species-parameters}(c), where $B_3$ is lowered before $B_5$. The outer barrier height $B_1$ is kept constant in
analogy with triple wells.

In order to limit the middle-well population, we impose two dynamic conditions at every
instant: (i) First, we keep dark state energy equidistant from the energies
of the neighboring states $\Delta^2E=\Delta E_e-\Delta E_g=0$. It can be seen from
Eq.~(\ref{adiabfactor}) that this reduces
transitions out of the dark state.  (ii) Secondly, we maintain the node of
the dark state at the center ($x_0$) of the middle well, $\left\langle x_0 | d\right\rangle = 0$. Geometrical considerations show that this lowers the middle-well population due to the antisymmetric shape of the wavefunction at the node.
Both of these conditions can be implemented with appropriate variation of the outer well depths $B_2$ and $B_6$ \cite{Opatrny-Das-PRA2009}, therefore in our simulations, it was convenient to directly reparameterize the lattice parameters $\{P_i,i\in[1,5]\}$ in terms of the parameter set  $\{B_1, B_3, B_5, \Delta^2E,  \left\langle x_0 | d\right\rangle \}$ that define the dark state, rather than using the intermediate set $\{B_i,i\in[1,2,3,5,6]\}$. Implementing the two dynamic conditions then simply requires setting $\Delta^2E=0$ and $\left\langle x_0 | d\right\rangle = 0$ for all points on the path.

\subsection{Compromising on Initial Conditions}
While the two dynamic
conditions set the relations between the parameters, the actual
values are set by initial conditions defined by our tolerances for the optimized physical features. The
key point is to recognize that all the undesirable
features are maximum at the halfway point of a
cycle (Fig.~\ref{1-species-parameters}(d)) when each cell is
\emph{symmetric} about a vertical line at its center. So, parameters
chosen to satisfy tolerances there will do so over the entire path. In this symmetric
configuration, three of the five parameters are used to maintain
the dynamic conditions: (i) dark state energy equidistant from the
neighboring energies, (ii) dark state node centered in the
middle well and (iii) the dark state exactly antisymmetric
with equal population in the extreme wells; the last two correspond to setting relative phases $P_4=P_5=0$.

There are \emph{two} free parameters remaining; however there are
at least four additional features that we would like to optimize:
(iv) maximum tolerance on central well population, $\alpha_m$ (v) the maximum
tolerance on the rate of leakage out of each cell, (vi) robustness
against noise and (vii) magnitude of separation $\Delta E (=\Delta
E_e=\Delta E_g$, both being equal in our scheme) between the
energy levels. These four factors are inter-related and obviously
all cannot be optimized with two parameters; so a right balance needs
to be found as we now discuss.

\subsection{Leakage}
One key distinguishing factor that differentiates lattice case from CTAP in a stand-alone triple well, is that the outer barriers of each unit cell cannot be made arbitrarily high due to the interplay of all the lattice parameters in defining the barrier heights and well-depths. This leads to finite rate of dissipation or leakage out of atom density out of each cell into adjacent cells. We find that a good measure of the tunneling or
leakage out of a cell is given by the width of the Bloch band
for the dark state $\Gamma_d =E_{d}(q=0)-E_{d}(q=2\pi/3\lambda)$, where $q$
is the Bloch vector for the lattice. This is primarily controlled by setting
the height of the outer barrier $B_1$ between unit cells. The relation of $\Gamma_d$ with $B_1$, although monotonically
decreasing, is not linear.

\begin{figure}[b]
\includegraphics*[width=\columnwidth]{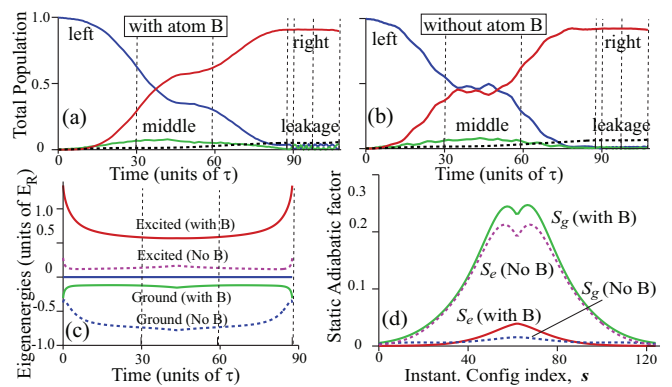}
\caption{(Color online)LCTAP with two species corresponding to Fig.~\ref{2-species-snapshots}, shown for one unit
cell and one cycle: Evolution of the population density with parameters \emph{optimized for presence of B} in the middle well
 with atom B (a) present and (b) absent. (c)
Evolution of the energies of the three states with and
without atom B (the dark state energies, \emph{different} in magnitude in
the two cases is used as reference for both for easy
comparison) (d) Evolution of the rate-independent static adiabatic
factors for the ground and excited states,  with and
without atom B, as a function of index labeling the instantaneous
states. }\label{2-species-parameters}
\end{figure}


\subsection{Noise Sensitivity}
We cannot predict the exact character of the imperfections of superlattices in actual experiments and the precise effect they would have on LCTAP evolution, until actual experiments are done. But here, we present some basic estimates of the tolerance of the process on noise and imperfections inevitable in experiments. Fast fluctuations in lattice parameters $P_i$ can induce non-coherent non-adiabatic coupling between eigenstates, which leads to a loss of population from the dark state to the neighboring states. The mean non-adiabaticity (there is no sum over $i$)
\bn
\beta_i \frac{dP_i}{dt} = \frac{1}{2} \sum_{j=e,g}
\frac{|\langle
\psi_j|\textstyle
  \frac{\partial V}{\partial P_i} | \psi_d \rangle|}
  {(E_j - E_d)^2}\frac{dP_i}{d\nu_i}\frac{d\nu_i}{dt}
\en
induced by any unwanted fluctuations $\nu_i$ in each of the time-varying lattice parameters $P_i$ is captured by the coefficient $\beta_i$ and depends on the rate of those fluctuations $\frac{d\nu_i}{dt}$. We found that by far, the dominant source of noise induced coupling between eigenstates arises from $\beta_4$ and $\beta_5$, associated with the relative phase parameters of the lattice potential. Plot of $\beta_4$ in Fig.~\ref{3d-plot}(a) shows
that effect of noise is reduced by parameter values that lead to higher leakage ($\Gamma_d$) and higher
middle well population ($\alpha_m$) both of which are undesirable, thereby indicating a compromise in the choice of parameters.

Our analysis is not dependent on the specific noise model.  We have chosen $1/f$ noise to probe the influence of generic noise in our model, since it is among the most ubiquitous in physical systems. An explanation of the widespread occurrence of such noise based on the self-organized criticality can be found in Ref.~\cite{Bak-PRL-1987}.  To all the parameters $P_i$, we therefore added the frequency-dependent noise  $\nu\propto\left[{\textstyle\sum_{n=1}^M} (f_n)^{-1}\cos(2\pi f_n t+\phi_n)\right]$ with randomly generated frequencies $f_n$ and phases $\phi_n$, which simulates technical noise with spectral density $1/f$.  We found that about $3\%$ noise ($\frac{dP_i}{d\nu_i}=0.03$ for \emph{all} $i$) still leads to successful LCTAP, although the transfer profile is somewhat distorted.

\subsection{Energy-level Separation}
\begin{figure}[t]
\includegraphics*[width=\columnwidth]{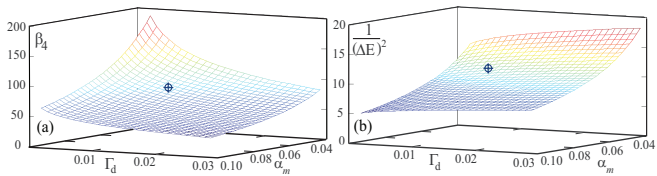}
\vspace{-4mm} \caption{(Color online) (a)
Coefficient of noise sensitivity$\beta_4$,
associated with relative phase $P_4$, plotted versus middle well
population $\alpha_m$ and the width of the dark-state Bloch band $\Gamma_d$ (measure of leakage rate out of unit cell); (b) Plot of inverse square of energy gaps between the dark and
adjacent states.
Parameters used in our simulations are marked in both; their locations in
the middle of the surfaces indicate trade-offs necessary in
optimization. Here $\Gamma_d$ and $\Delta E$ are in units of $E_R$.}\label{3d-plot}
\end{figure}

Among all the parameters involved in LCTAP, the most important one is the energy
separation $\Delta E$ between the dark state and its adjacent states, since it is related to almost all the optimization
factors, and  Eq.~(\ref{adiabfactor}) suggests larger
$\Delta E$ would be better. But this
corresponds to lowering the internal barriers $B_3$ and $B_5$
which leads to a higher slope of the dark state in the neighborhood of the node, leading to higher
occupation of the middle well. This is a static effect defined by the shape of the \emph{instantaneous} eigenstates, not to be
confused with effects of transitions induced by nonadiabticity -
which do benefit from a higher energy separation of the states. So a
compromise has to be struck; Figs.~\ref{3d-plot} show that if noise and
energy separation are to be kept optimal as well, we need to
sacrifice on tunneling and maximum middle well population. We chose our parameters by minimizing the cost of
trade-offs among the optimized features.

\subsection{Time Evolution}
The simplest possibility
for the dynamical factor $ds/dt$ is to keep it constant, then the
tolerances are determined by the maximum of the instantaneous
factors ${\cal S}_j$ (Fig.~\ref{1-species-parameters}(d)). However, recognizing that it is the product  $ {\cal A}_j={\cal
S}_j\frac{ds}{dt}$ that needs to be within tolerances, LCTAP can be speeded up significantly, by keeping ${\cal A}_j$ fixed and dynamically adjusting the rate
$ds/dt={\rm{max}({\cal A}_e,{{\cal A}_g}})/{{\rm max}({{\cal
S}_e,{\cal S}_g}})$; this can double the cycle speed,
significant in a coherent mechanism.

\subsection{Dual Species}
All of our analysis applies when species B is present in the middle well, but with optimization done for the modified  potential felt by A. However, to use the \emph{same} LCTAP \emph{parameter} evolution for \emph{both with and without} the atom B present, we change the energy balance condition to
$\Delta^2E=\Delta E_e(without\ species\ B)-\Delta E_g(with\
species \ B) =0$, which takes the two most proximate levels (as seen in
Fig.~\ref{2-species-parameters}(c)) and maintains them equally apart from the dark state. The dynamic transfer rate $ds/dt$ is now set by  the maximum of \emph{four} adiabatic factors (Fig.~\ref{2-species-parameters}(d)) corresponding to the ground and excited states,
with and without the atoms B.

\section{Physical Parameters and Feasibility}
\label{sec-Parameters}
We now put our
parameters in the context of recent experiments with $^{87}$Rb
in superlattices, choosing laser wavelengths around $803$
nm close to the D1 and D2 lines. This translates to time
unit $\tau=45$ microseconds putting our LCTAP cycles
at $2.5-4.5$ milliseconds near accessible time scales
\cite{Phillips-PRL2007}.
Figures \ref{1-species-snapshots} and \ref{2-species-snapshots} show
that the individual lattice depths ($P_1,P_2,P_3$ in units of
$4E_R$) for the target species A range between $2E_R-8 E_R$, their
sum never exceeding about $10 E_R$. We found that to keep the
species B unaffected as target species A evolves, $V_B/V_A\simeq6$
is sufficient \cite{movies} implying a maximum net lattice depth for species B of about $60
E_R$, accessible in current experiments \cite{Phillips-PRL2007,Phillips-Nature2007}. Figures \ref{1-species-parameters} and \ref{2-species-parameters} use
a higher value $V_B/V_A=20$ only to show the interaction potential as an obvious `bump'.

The ground state density of atom B is a Gaussian of
$\sigma=0.035 \lambda$ and peak density $|\psi^B_0(0)|^2=11.4/\lambda$. For our estimates, we assume two hyperfine states \cite{Dominik-Schneble-PRL2010} of $^{87}$Rb and use its triplet scattering length $a=99a_B$: For a uniform transverse lattice confinement equal to $V_B=60E_R$ we get $\omega_\perp=(2\pi/\lambda)\sqrt{2V_B/m}=3.4\times 10^5$ Hz, $g_{1D}=0.20 E_R\lambda$ and barrier height $g_{1D}|\psi^B_0(0)|^2=2.3E_R$. It can be much less for a transverse magnetic confinement on an atom chip \cite{Das-Aubin-PRL2009}, $\omega_\perp\simeq2\pi\times 5100$ Hz and $g_{1D}=0.019E_R\lambda$. Our simulations use a cautious higher value for the barrier height $0.30|\psi^B_0(0)|^2=3.4 E_R$.

Finally, we note that high transfer efficiency was achievable in our simulations. In the case of single species, shown in Figs.~\ref{1-species-snapshots} and \ref{1-species-parameters}, at the end of the first cycle $94.6\%$ of the population ended up in the right well with $4.7\%$ overall lose out of each cell, the remainder distributed between the first and the middle well. With dual species, as shown in Figs.~\ref{2-species-snapshots} and \ref{2-species-parameters}, the corresponding numbers were $90.8\%$ and $6.3\%$ with species B present in the central well; and $91.3\%$ and $7.5\%$ with species B absent in the central well. These values indicate that multiple cycles can be repeated in a single run of an experiment.

\section{Conclusion}
\label{Conclusion}
We note that our use of a three-mode localized state is validated by our numerical results: (i) there is low leakage of density out of each cell per cycle and (ii) the width of the dark state Bloch band $\Gamma_d\sim 0.01 E_R$ (Fig.~\ref{3d-plot}) is an order of
magnitude less than the energy separations (Fig.~\ref{1-species-parameters} and \ref{2-species-parameters}). Our results indicate viable implementation of LCTAP and should motivate experiments with triple-well optical lattices. By allowing choice of the degree and the time of overlap of the two species this mechanism can be tailored for controlled interaction-induced change of internal states.

\begin{acknowledgments}
  T.O. and M.G. acknowledge support of the Czech Science Foundation, grant P205/10/1657. K.K.D. acknowledges the support of the NSF under grant PHY-0970012. We also gratefully for valuable conversations with Dominik Schneble and Seth Aubin.
\end{acknowledgments}

\appendix
\section{Numerical reparametrization}
\label{appA}

A key step in our simulations is the reparameterization of the lattice variables $\{P_i,i\in[1,5]\}$ of amplitude and phase in terms of characteristics of the localized states of individual triple-well unit cells.

The lattice Hamiltonian $H\equiv H(P_i(t))$ is characterized by five time-dependent lattice parameters, $\{P_i,i\in[1,5]\}$. We identify five characteristics  $\{F_j,j\in[1,5]\}$  of the Hamiltonian  that need to be optimized for LCTAP. The Hamiltonian can be numerically reparametrized in terms of these properties $F_j(t) \equiv F_j[H(P_i(t))]=F_j[P_i(t)]$ provided that the Jacobian matrix $J_{ji} = \frac{\partial F_j}{\partial P_i}$ has nonzero determinant in the region of the reparametrization.

The properties $F_j$ are our operational parameters, with the advantage that our optimization conditions for LCTAP directly determine the dynamic values of these new parameters $F_j$. In our numerical simulations, where the evolution is done by split step operator method, at every time step $t_n$ of the evolution we therefore have a well defined set of \emph{known optimal} parameters $F_j(t_n)$. In order to determine how the lattice needs to evolve for optimal LCTAP, at each of these steps we need to determine the corresponding optimal lattice parameters $P_i(t_n)$. We do this iteratively as follows:

At \emph{every} time step $t_n$, start with a reasonable guess $P_i^{(1)}$ for the lattice parameters  based upon the values of the previous time step, but which correspond to \emph{non-optimal} values $F_j^{(1)}=F_j(P_i^{(1)})$. Using the differences $\delta F_j^{(1)} = F_j(t_n) - F_j^{(1)}$ in the system of linear equations $\delta F_j^{(1)} = \sum_i \frac{\partial F_j}{\partial P_i} \delta P_i^{(1)} = \sum_i J_{ji} \delta P_i^{(1)}$, we solve for $\delta P_i^{(1)}$ by inverting the Jacobian matrix, and thereby find a new improved set of parameters $P_i^{(2)}=P_i^{(1)}+\delta P_i^{(1)}$ with the new values $F_j^{(2)}=F_j(P_i^{(2)})$ \emph{closer} to the optimal values $F_j(t_n)$. We repeat this for $k$ iterations until $\delta F_j^{(k)} = F_j(t_n) - F_j^{(k)}$ is less than our set tolerance.

As described in Sec.~\ref{sec-Analysis}, the appropriate properties $F_j$ are chosen to be $\{B_1, B_3, B_5, \Delta^2 E, \left\langle x_0 | d \right\rangle \}$.  As discussed in Sec.~\ref{sec-Analysis}, we set $\Delta ^2 E = 0$ and $\left\langle x_0 | d \right\rangle = 0$ for all points of the path to meet our conditions for LCTAP.

\end{document}